\documentclass{optica-article}

\journal{opticajournal} 

\newcommand{\pos}{\mathbf{r}}
\newcommand{\dir}{\hat{\mathbf{s}}}
\newcommand{\inc}{\hat{\mathbf{\imath}}}
\newcommand{\obs}{\hat{o}}

\begin{document}

\title{Scattering by nanoplasmonic mesoscale assemblies}

\author{Md. Imran Khan,\authormark{1,*}, Sayantani Ghosh\authormark{1},
  and Arnold D.~Kim,\authormark{2}}

\address{Department of Physics\authormark{1} and Applied
  Mathematics\authormark{2}, University of
  California, Merced, 5200 North Lake Road, Merced, CA 95343, USA}

\email{\authormark{*}mkhan39@ucmerced.edu}

\begin{abstract*}
  The flexibility and versatility of nanoassembled plasmonic structures provide platforms for mesoscale tunable optical modulation. Our recently developed model for these nanoassembled plasmonic structures is composed of a dielectric spherical core surrounded by a concentric spherical shell containing a random
distribution of AuNPs. This model provides a useful platform for studying the role of a controlled amount of disorder on scattering by a particle. In that context, we explore the angular distribution of scattered light for different sizes (5 - 20 nm) and filling fractions (0.1 - 0.3) of the AuNP in the coatings. The simulations reveal that the coating of AuNPs
  redistributes power in a way that suppresses angular side lobes,
  thereby guiding the scattered power preferentially in the forward
  direction. These results highlight that with the ability to tune both the spatial and the spectral aspects of the scattering profile,
  these coated structures may serve as a platform for a variety of applications, including passive cloaking, scattering enhancement, and high-resolution
  imaging.
\end{abstract*}

\section{Introduction}

Because they exhibit localized surface plasmonic
coupling~\cite{lsprexploitation, lspr_spectroscooy, Mayer2011,
  lspr_biosensors, PETRYAYEVA20118}, metal nanoparticles have been
used to control the properties of electromagnetic waves,
opening opportunities for photonics applications~\cite{alu_2017,
  Ross2016, Liu2008, Garcia-Lojo2019}. Some of the unique
functionalities include materials with extraordinary transmission,
optical magnetism, and photonic lensing~\cite{zhu2019, McCloskey:12,
  Choo2012, Gordon2008, Im2014, Ebbesen1998, Linden2006}. As in all
cases of nanoassembly, these meta-structures can be constructed using
either top-down or bottom-up methods. Bottom-up methods offer greater
flexibility in terms of composition and morphology, while also
providing a route to generate large-scale structures, extending
from the nano-scale to the mesoscale and beyond~\cite{Cunningham, Mayer2019,
  Ghosh2021}. Colloidal synthesis techniques, such as directed assembly modulated with polymer templates, DNA-mediated, or liquid crystals,
are some of the few methods that have produced meta-structures composed of metal nanoparticles that have demonstrated tailored
permittivity, photoconductivity, local surface plasmonic tuning, and
passive cloaking through scattering suppression~\cite{Cunningham,
  Mayer2019, Ross2015,Young2014, Rodarte2013, Chen2012, Rodarte2015}.

The authors have introduced a model for nano-assembled plasmonic
metastructures consisting of a dielectric spherical core surrounded by
concentric spherical shell of randomly distributed metal
nanoparticles~\cite{Khan}. Rather than resorting to using effective
medium theory or anomalous diffraction theory to model the interaction
between the core and shell which have inherent scale
limitations~\cite{Cunningham, Ross2015, Choy2015, MASLOWSKA199435},
this model explicitly incorporates multiple interactions between
individual metal nanoparticles in the shell and the core, allowing for
core sizes comparable to the wavelength of incident
light. Consequently, this model provides valuable insight into the
optical properties of these metastructures across a broad range of
spatial and spectral scales. We have used this model to investigate
how key design parameters (nanoparticle size, filling fraction in the
shell, etc.)  affected the extinction properties of nano-assembled
plasmonic metastructures~\cite{Khan}. More specifically, we have shown
that this model exhibits a wide variety of complex behaviors between enhancement and suppression of scattering when core sizes are comparable
to the wavelength~\cite{Khan}, which is precisely a scale not
accurately modeled using effective medium theory.

The model couples scattering by a dielectric core with the multiple
scattering by the ensemble of individual metal nanoparticles that are
randomly distributed in the shell. In doing so, this model provides a
theoretical framework for investigating how a controlled amount of
disorder introduced by the metal nanoparticles affects the scattering by a
particle. Even though aspects of this problem are physically
intuitive, this model provides a systematic and quantitative method to
explore fundamental mechanisms in these problems which, in turn, may
lead to valuable insights for photonics applications. Our objective
here is to go beyond just extinction properties and evaluate the
elementary scattering properties of nanoassembled plasmonic
metastructures to identify how the introduction of disordered metal
nanoparticles affects scattering. Consider the classical Mie scattering problem in which a monochromatic
plane wave is incident on a dielectric
sphere~\cite{bohren2008absorption}. The amplitude and phase of the field scattered by the sphere is the result of refraction and
diffraction due to the boundary surface and interior material of the
sphere~\cite{akira_ishimaru}. Mathematically, the Mie solution is
computed by interchanging between a cartesian coordinate system
natural for the incident field to a spherical coordinate system
natural for the scattered field. The cumulative result leads to
complex behavior of the scattered field, especially when the sphere is
comparable to the wavelength of the incident field. We typically
expect that the monochromatic field scattered by a dielectric sphere
whose size is comparable to the wavelength to exhibit a primary
forward scattering peak, a secondary backscattering peak, and
oscillatory lobes in other directions.
\begin{figure}[htb]
    \centering
    \includegraphics[width = 6 cm]{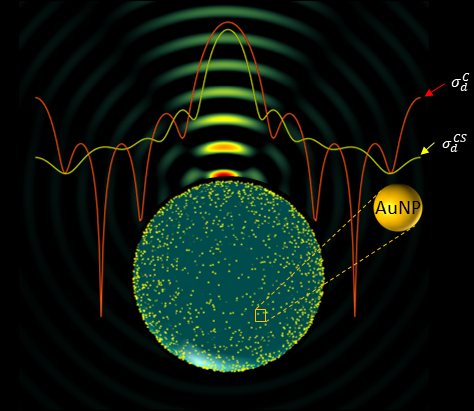}
    \caption{The core-shell structure represented by a blue sphere
      (core) coated with plasmonic AuNPs (shell). The red pattern
      shows the differential scattering cross-section of the uncoated
      dielectric sphere ($\sigma_d ^C $), while the yellow pattern
      represents the same dielectric core coated with AuNPs
      ($\sigma_d^{CS}$). The differential scattering cross-section of
      a plasmonic composite structure demonstrates a suppressed
      central lobe and smoother angular side lobes compared to the
      corresponding bare dielectric sphere for any particular incident
      wavelength.}
    \label{fig:fig_1}
\end{figure}
Now consider a nano-assembled plasmonic meta structure consisting of a
spherical dielectric core surrounded by a concentric shell containing
a disordered ensemble of gold nanoparticles (AuNPs) as shown in
Fig.~\ref{fig:fig_1}. Each AuNP absorbs and scatters depending on its
size and material composition. The ensemble of AuNPs in the shell will
act as a perturbation of Mie scattering by the dielectric spherical
core. The overall strength of that perturbation will depend strongly
on the scattering properties of the AuNPs and their filling fraction
in the shell. Multiple scattering by the AuNPs will strongly perturb
the phase of the scattered field by the core because the disorder
introduces a decoherence mechanism that interferes with the
aforementioned refraction and diffraction by the dielectric sphere.
Although we have seen that the cumulative effect of the shell of AuNPs
can drastically alter the extinction of the core, we seek to
understand more details about the scattering properties of these meta structures. The remainder of this paper is as follows. In Section \ref{sec:model}
we review the features and details of the computational model. In
Section \ref{sec:scattering} we review the theory of scattering by a particle and how to compute the elementary cross-sections from the
results of the computational model. We give our computational results
in Section \ref{sec:results} that identify the key changes in behavior
of scattering due to the ensemble of metal nanoparticles in the
shell. Section \ref{sec:conclusions} contains our conclusions.
\section{The model}
\label{sec:model}

To model scattering by a spherical core surrounded by a concentric
shell of randomly distributed AuNPs, we consider several fields: the
interior field inside the core, denoted by $\psi^{\text{int}}$, the
field scattered by the core, denoted by $\psi^{\text{core}}$, and the
field scattered by each of the $N$ AuNPs, denoted by
$\Psi_{n}^{\text{NP}}$ for $n = 1, \dots, N$. We describe our method
for computing each of these fields below.

\subsection{Interior and exterior fields}

Let the core be a sphere of radius $a$ centered at the origin of our
coordinate system so that $| \pos | < a$ corresponds to its interior,
$| \pos | > a$ corresponds to its exterior, and $| \pos | = a$
corresponds to its boundary surface. The wavenumber interior to the
core is $k_{1}$ and exterior to the core is $k_{0}$.

We use the method of fundamental solutions (MFS) to compute
$\psi^{\text{int}}$ and $\psi^{\text{core}}$. To do so, let $\dir_{m}$
for $m = 1, \dots, M$ denote $M$ points that sample the unit sphere so
$| \dir_{m} | = 1$ for $m = 1, \dots, M$. We introduce a length
$\ell \ll a$ and write
\begin{equation}
  \psi^{\text{int}}(\pos) \approx \sum_{m = 1}^{M} 
  \frac{e^{\mathrm{i} k_{1} ( \pos - ( a + \ell )
      \dir_{m} )}}{4 \pi ( \pos - (a + \ell)
    \dir_{m} )} c_{m}, \quad | \pos | < a,
  \label{eq:interior}
\end{equation}
This expression gives $\psi^{\text{int}}$ as a superposition of
Green's functions. The expansion coefficients $c_{m}$ for
$m = 1, \dots, M$ are to be determined. Since
$| ( a + \ell ) \dir_{m} | > a$ for $m = 1, \dots, M$, all the source
points lie in the exterior. Therefore, the evaluation of
\eqref{eq:interior} is an {\em exact} solution of
\begin{equation}
  ( \nabla^{2} + k_{1}^{2} ) \psi^{\text{int}} = 0, \quad | \pos | < a.
\end{equation}
Similarly, we write
\begin{equation}
  \psi^{\text{core}}(\pos) \approx \sum_{m = 1}^{M}
  \frac{e^{\mathrm{i} k_{0} ( \pos - ( a - \ell ) \dir_{m})}}{4 \pi (
    \pos - (a - \ell) \dir_{m})}  b_{m}, \quad | \pos | > a.
  \label{eq:exterior}
\end{equation}
Here, the expansion coefficients $b_{m}$ for $m = 1, \dots, M$ are to
be determined.  In \eqref{eq:exterior} the source points all satisfy
$| ( a - \ell ) \dir_{m} | < a$ for $m = 1, \dots, M$, so they all lie
in the interior. Consequently, this superposition of Green's functions
{\em exactly} satisfies
\begin{equation}
  ( \nabla^{2} + k_{0}^{2} ) \psi^{\text{int}} = 0, \quad | \pos | > a.
\end{equation}
Moreover, each Green's function satisfies the correct radiation
condition for $|\pos| \to \infty$, so \eqref{eq:exterior} ensures that
that $\psi^{\text{core}}$ is appropriately outgoing.

\subsection{Scattered fields by gold nanoparticles}

Assuming that the AuNPs are small compared to the wavelengths in the visible spectrum, we model them as isotropic point scatterers:
\begin{equation}
  \Psi^{\text{NP}}_{n}(\pos) = \alpha_{n} \frac{e^{\mathrm{i} k_{0} |
      \pos - \pos_{n}|}}{4 \pi | \pos - \pos_{n} |} \psi^{E}_{n},
  \quad n = 1, \dots, N,
  \label{eq:NP-scattered}
\end{equation}
with $\pos_{n}$ denoting the center position of the $n$th gold
nanoparticle, $\alpha_{n}$ denoting its scattering amplitude, and
$\psi_{n}^{E}$ denoting the exciting field. Let $\sigma_{s}$ and
$\sigma_{t}$ denote scattering and total cross-sections for an
individual AuNP. The scattering amplitude $\alpha_{n}$ is given
by~\cite{Khan}
\begin{equation}
  \alpha = \left[ \frac{\sigma_{s}}{4\pi} - \left( \frac{k_{0}
        \sigma_{t}}{4 \pi} \right)^{2} \right]^{1/2} + \mathrm{i}
  \frac{k \sigma_{t}}{4 \pi}.
\end{equation}

\subsection{Boundary conditions}

We determine the expansion coefficients, $b_{m}$ and $c_{m}$ for
$m = 1, \dots, M$, and exciting fields $\psi^{E}_{n}$ for
$n = 1, \dots, N$ using a collocation method for boundary conditions
on the surface of the sphere that we explain below.
The field scattered by the nanoplasmonic mesoscale assembly is
\begin{equation}
  \psi^{\text{sca}} = \psi^{\text{core}} + \sum_{n = 1}^{N}
  \Psi_{n}^{\text{NP}}.
  \label{eq:scattered-field}
\end{equation}
We relate $\psi^{\text{sca}}$ with $\psi^{\text{int}}$ and the
incident field $\psi^{\text{inc}}$ through the following conditions on
the boundary of the core:
\begin{equation}
  \psi^{\text{inc}} + \psi^{\text{sca}} = \psi^{\text{int}}  \quad
  \text{and} \quad
  \partial_{n} \psi^{\text{inc}} + \partial_{n} \psi^{\text{sca}} =
  \partial_{n} \psi^{\text{int}} \quad \text{on $|\pos| = a$}.
  \label{eq:BCs}
\end{equation}
Note that different boundary conditions can be specified here without
any additional difficulty in implementation.

For this collocation method, we require that the conditions in
\eqref{eq:BCs} are satisfied on each of the points
$\pos^{\text{bdy}}_{m} = a \dir_{m}$ for $m = 1, \dots,
M$. Substituting \eqref{eq:exterior} and \eqref{eq:NP-scattered} into
\eqref{eq:scattered-field} and substituting that result along with
\eqref{eq:interior} into the first condition in \eqref{eq:BCs}
evaluated at these boundary points yields,
\begin{multline}
  \psi^{\text{inc}}(a \dir_{m}) + \sum_{m' = 1}^{M} G_{0}( | a
  \dir_{m} - ( a - \ell ) \dir_{m'} | ) b_{m'} + \sum_{n = 1}^{N}
  \alpha_{n} G_{0}( | a \dir_{m} - \pos_{n}| ) \psi^{E}_{n}\\
  = \sum_{m' = 1}^{M} G_{1}( | a \dir_{m} - ( a + \ell )
  \dir_{m'} | )  c_{m'}, \quad m = 1, \dots, M,
  \label{eq:BC1}
\end{multline}
with $G_{0}(R) = e^{\mathrm{i} k_{0} R}/(4\pi R)$ and $G_{1}(R) =
e^{\mathrm{i} k_{1} R}/(4\pi R)$. Performing the same procedure for
the second condition in \eqref{eq:BCs} and using
$\partial_{n} = \partial_{r}$ for a sphere, we obtain
\begin{multline}
  \partial_{r} \psi^{\text{inc}}(a \dir_{m}) + \sum_{m' = 1}^{M}
  \partial_{r} G_{0}( | a \dir_{m} - ( a - \ell ) \dir_{m'} | ) b_{m'}
  + \sum_{n = 1}^{N} \alpha_{n} \partial_{r} G_{0}( | a \dir_{m} -
  \pos_{n}| ) \psi^{E}_{n}\\
  = \sum_{m' = 1}^{M} \partial_{r} G_{1}( | a \dir_{m} - ( a + \ell )
  \dir_{m'} | ) c_{m'} , \quad m = 1, \dots, M.
  \label{eq:BC2}
\end{multline}

\subsection{Exciting fields}

To determine the exciting fields $\psi^{E}_{n}$ for $n = 1, \dots, N$,
we use the self-consistent Foldy-Lax theory~\cite{foldy,
  lax}. Consider the $n$th AuNP. The exciting field $\psi^{E}_{n}$ is
given by the fields scattered by the core and all {\em other}
AuNPs. Using our approximations, this definition of the exciting
fields yields
\begin{equation}
  \psi^{E}_{n} = \sum_{m = 1}^{M} G_{0}( | \pos_{n} - ( a - \ell
  ) \dir_{m} | ) b_{m} + \sum_{\substack{n' = 1\\n' \neq n}}^{N} \alpha_{n'}
  G_{0}( | \pos_{n} - \pos_{n'} | ) \psi^{E}_{n'}, \quad n = 1, \dots,
  N.
  \label{eq:exciting-fields}
\end{equation}

\subsection{Block system}

Equations \eqref{eq:BC1}, \eqref{eq:BC2}, and
\eqref{eq:exciting-fields} give $2M + N$ equations for the $2M$
expansion coefficients $b_{m}$ and $c_{m}$ for $m = 1, \dots, M$, and
$N$ exciting fields $\psi^{E}_{n}$ for $n = 1, \dots, N$. Let
$\mathbf{b} = (b_{1}, \dots, b_{M})$,
$\mathbf{c} = (c_{1}, \dots, c_{M})$ and
$\boldsymbol{\psi}^{E} = (\psi^{E}_{1}, \dots,
\psi^{E}_{N})$. Equations \eqref{eq:BC1}, \eqref{eq:BC2}, and
\eqref{eq:exciting-fields} can be combined leading to the following block system:
\begin{equation}
  \begin{pmatrix}
    A_{1} & A_{2} & B_{1} \\
    A_{3} & A_{4} & B_{2} \\
          & C    & D
  \end{pmatrix}
  \begin{pmatrix}
    \mathbf{b}\\ \mathbf{c} \\ \boldsymbol{\psi}^{E}
  \end{pmatrix}
  =
  \begin{pmatrix}
    \mathbf{f} \\ \mathbf{g}  \\ 0
  \end{pmatrix}.
  \label{eq:block-system}
\end{equation} 
Here, the $M \times M$ blocks $A_{1}$, $A_{2}$, $A_{3}$ and $A_{4}$ have entries
\begin{align*}
  [ A_{1} ]_{m,m'} &= - G_{0}( | a \dir_{m} - ( a - \ell ) \dir_{m'} | ),\\
  [ A_{2} ]_{m,m'} &= G_{1}( | a \dir_{m} - ( a + \ell ) \dir_{m'} |
                     ),\\
  [ A_{3} ]_{m,m'} &= - \partial_{r} G_{0}( | a \dir_{m} - ( a - \ell
                     ) \dir_{m'} | ),\\
  [ A_{4} ]_{m,m'} &= \partial_{r} G_{1}( | a \dir_{m} - ( a + \ell )
                     \dir_{m'} | ).
\end{align*}
The $M \times N$ blocks $B_{1}$ and $B_{2}$ have entries
\begin{align*}
  [ B_{1} ]_{m,n} &= -G_{0}( | a \dir_{m} - \pos_{n} | ) \alpha_{n},\\
  [ B_{2} ]_{m,n} &= -\partial_{r} G_{0}( | a \dir_{m} - \pos_{n} | )
                    \alpha_{n}.
\end{align*}
The $N \times M$ block $C$ has entries
\begin{equation*}
  [ C ]_{n,m} = -G_{0}( | \pos_{n} - ( a - \ell ) \dir_{m} | ),
\end{equation*}
and the $N \times N$ block $D$ has entries
\begin{equation*}
  [ D ]_{n,n'} = \begin{cases}
    1 & n = n',\\
    -G_{0}(| \pos_{n} - \pos_{n'} |) \alpha_{n'} & n \neq n'.
  \end{cases}.
\end{equation*}
The block vectors $\mathbf{f}$ and $\mathbf{g}$ have entries
\begin{align*}
  [ \mathbf{f} ]_{m} &= \psi^{\text{inc}}(a \dir_{m}),\\
  [ \mathbf{g} ]_{m} &= \partial_{r} \psi^{\text{inc}}(a \dir_{m}).
\end{align*}
Upon solution of \eqref{eq:block-system}, we can compute
$\psi^{\text{int}}$ through evaluation of \eqref{eq:interior},
$\psi^{\text{core}}$ through evaluation of \eqref{eq:exterior}, and
$\Psi^{\text{NP}}_{n}$ through evaluation of equation
\eqref{eq:NP-scattered}. With $\psi^{\text{core}}$ and
$\Psi^{\text{NP}}_{n}$ for $n = 1, \dots, N$ determined, we can
evaluate \eqref{eq:scattered-field} and study its properties which we
discuss below.

\section{Scattering properties}
\label{sec:scattering}

Suppose the incident field is a plane wave with unit amplitude
propagating in direction $\hat{\mathbf{\imath}}$,
$\psi^{\text{inc}}(\pos) = e^{\mathrm{i} k_{0} \inc \cdot
  \pos}$. Using that incident field in \eqref{eq:block-system}, we
compute the far-field behavior of $\psi^{\text{sca}}$ given in
\eqref{eq:scattered-field} by making use of the asymptotic behavior of
Green's function,
\begin{equation}
  G_{0}(| R \hat{\mathbf{o}} - \pos' |) \sim e^{-\mathrm{i} k_{0}
    \hat{\obs} \cdot \pos'} \frac{e^{\mathrm{i} k_{0}
      R}}{4 \pi  R }, \quad R \gg 1.
\end{equation}
When we replace $G_{0}$ by this far-field asymptotic behavior in
\eqref{eq:scattered-field}, we obtain
\begin{equation}
  \psi^{\text{sca}}(R \hat{\mathbf{o}}) \sim \left[
    \frac{1}{4 \pi} \sum_{m = 1}^{M} e^{- \mathrm{i} k_{0} ( a - \ell )
        \obs \cdot \dir_{m}} b_{m}(\inc) + \frac{1}{4 \pi} \sum_{n =
        1}^{N} \alpha_{n} e^{-\mathrm{i}
        k_{0} \obs \cdot \pos_{n}} \psi^{E}_{n}(\inc) \right]
  \frac{e^{\mathrm{i} k_{0} R}}{R}, \quad R \gg 1.
\end{equation}
Note that we have added the explicit dependence on the incident
direction $\inc$ in the expression above.  We call
\begin{equation}
  f(\obs,\inc) = \frac{1}{4 \pi} \sum_{m = 1}^{M} e^{- \mathrm{i}
    k_{0} ( a - \ell ) \obs \cdot \dir_{m}} b_{m}(\inc) + \frac{1}{4
    \pi} \sum_{n = 1}^{N} \alpha_{n} e^{-\mathrm{i} k_{0} \obs \cdot
    \pos_{n}} \psi^{E}_{n}(\inc),
\end{equation}
the scattering amplitude for the nanoplasmonic mesoscale assembly. It
gives the amplitude and phase of the scattered field in the far-field
in direction $\obs$ due to an incident plane wave in direction $\inc$.

We use this scattering amplitude to compute cross-sections that are
used in studying scattering by particles~\cite{bohren2008absorption,
  akira_ishimaru}.  First, to characterize the distribution of power
scattered by the nanoplasmonic mesoscale assembly, we compute the
differential scattering cross-section,
\begin{equation}
  \sigma_{d}(\obs,\inc) = | f(\obs,\inc) |^{2}.
\end{equation}
Using the Optical Theorem, we compute the total cross-section
(extinction) according to
\begin{equation}
  \sigma_{t} = \frac{4 \pi}{k_{0}} \text{Im}\{ f(\inc,\inc) \}.
\end{equation}
The scattering cross-section is defined as
\begin{equation}
  \sigma_{s} = \int_{4\pi} \sigma_{d}(\obs,\inc) \mathrm{d}\obs.
\end{equation}

Using these cross-section, we introduce two non-dimensional parameters
used to characterize scattering.  The scattering albedo $\varpi_{0}$
is given as
\begin{equation}
  \varpi_{0} = \sigma_{s}/\sigma_{t},
\end{equation}
and gives the fractional amount of $\sigma_{t}$ due to
$\sigma_{s}$. When $\varpi_{0} = 1$, extinction is due entirely to
scattering and when $\varpi_{0} = 0$ it is due entirely to
absorption. The anisotropy factor $g$ is defined as
\begin{equation}
  g = \frac{1}{\sigma_{t}} \int_{4\pi} ( \obs \cdot \inc )
  \sigma_{d}(\obs,\inc) \mathrm{d}\obs.
\end{equation}
The anisotropy factor gives a measure for the amount of power flow in
forward/backward directions. When $g = 0$ scattering is isotropic and
when $g = \pm 1$ scattering is purely in the forward or backward
directions, respectively.

\section{Results}
\label{sec:results}

For the following results, we have set relative refractive index for
the spherical core to be $n_{r} = 1.4$ corresponding to silica. We
varied the core diameter from $450$ nm to $750$ nm, and considered the
AuNPs' diameter to be either $5$ nm, $10$ nm, or $20$ nm.  AuNPs have
ligands associated with them for surface functionalization~\cite{Khan}
and we have set the ligand length to $0.98$ nm.  The AuNPs in the
shell are distributed randomly and the minimum distance between AuNPs
and the silica core is $0.98$ nm since the ligand that maintains the
structure and surface functionalization of AuNPs also sustains this
separation distance. We use the optical properties of the AuNPs from experiments
conducted on plasmonic film developed for plasmonic and nanophotonic
applications by Yakubovsky
$\textit{et al}$~\cite{Yakubovsky:17}. The number $N$ of the AuNPs
that we used depends on the filling fraction ($f$) in each case as we
have explained in \cite{Khan}. To compute $\sigma_s$ and $g$, we used
the product Gauss quadrature rule~\cite{atkinson} with M = $512$
quadrature points.

\begin{figure}
    \centering
    \includegraphics[width = 6 cm]{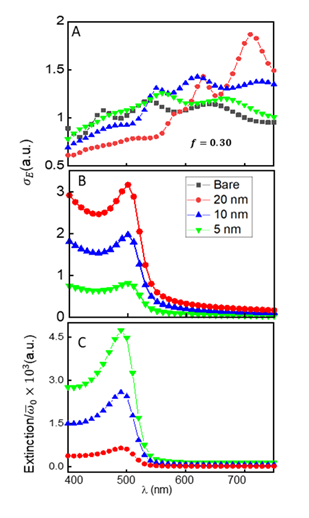}
    \caption{(A) $\sigma_E $ of a 750 nm bare silica sphere (black
      square) compared with the plasmonic core-shell structures
      comprised of the same core diameter but coated with varied AuNP
      diameters. The filling fraction was 0.30 for all AuNP sizes. A 20 nm
      (red dots) shell shows scattering suppression
      ($\sigma_E^{CS} < \sigma_E^S$) from 400 nm to 600 nm of the
      visible spectrum. A 10 nm (blue triangle) and 5 nm (green,
      inverted triangle) core-shell are barely able to suppress
      scattering ($\sigma_E^{CS} \approx \sigma^C$ ) from 400 nm to 550
      nm of the visible spectrum.  All core-shell structures show
      scattering enhancement ($\sigma_E^{CS} > \sigma_E^C $ ) beyond 600
      nm wavelength. (B) Extinction cross-section (scattering plus
      absorption) of a single AuNP over the visible spectrum. AuNP
      diameters varied as 5 nm, 10 nm, and 20 nm. (C) Ratio of
      extinction to their respective albedo. This ratio indicates absorption
      contributes more to the extinction as the AuNP size decreases.}
    \label{fig:2}
\end{figure}

We compute and plot the scattering efficiency, defined according
to
\begin{equation*}
  \sigma_E = \frac{\sigma_t(\lambda)}{\sigma_g}
\end{equation*}
as function of core diameter $d$ with $\sigma_g$ denoting the
geometric cross-section. Figure \ref{fig:2}(A), shows the scattering
efficiency for a dielectric core $\sigma_{E}^{C}$ and a dielectric core
surrounded by a plasmonic AuNP shell $\sigma_{E}^{CS}$. For the silica
core, $\sigma_g = \pi R_c^2$ with $ R_c=d/2$ denoting the radius of the core,
and $\sigma_g = \pi R_s^2$ for the core-shell assembly with $R_s$
denoting the shell outer radius.

In Fig.~\ref{fig:2}A, $\sigma_E^C$ exhibits oscillatory behavior
characteristic of the Mie theory~\cite{bohren2008absorption}.  The peaks
of those oscillations correspond to the so-called Mie resonances. The
locations and heights of these peaks are characteristic of the size
and relative refractive index of the sphere. In the case of plasmonic
core-shell assemblies, the diameter of the core is constant at $750$ nm, but the
AuNPs of the shell are either $5$ nm (green inverted triangle), $10$
nm (blue triangle) or $20$ nm (red circle). For all of these cases,
the AuNP filling fraction in the shell is held constant at $f =
0.30$. We observe that a shell with $20$ nm AuNPs suppresses
scattering ($\sigma_E^{CS} < \sigma_E^C$) from $400$ nm to $600$ nm of
the visible spectrum. Here the core-shell composite shows little to no
trace of Mie resonances associated with the silica core. On the other
hand, the shell with $5$ nm AuNPs barely suppresses scattering and
closely follows $\sigma_E^C$. The shell with $10$ nm AuNPs has a
slightly higher ability to suppress scattering compared to the $5$ nm
AuNPs but is considerably lower than the $20$ nm AuNPs.

Figure~\ref{fig:2}B shows $\sigma_{E}$ for individual AuNPs. Note that
the efficiencies for all three AuNP sizes exhibit an exponential decay
as $\lambda$ increases. Because of this property, we find that at
these larger wavelengths, the shell of AuNPs yields enhanced
scattering efficiencies ($\sigma_E^{CS} > \sigma_E^C$) shown in
Fig.~\ref{fig:2}A. The shell with $20$ nm AuNPs has the highest
extinction, especially for $\lambda > 600$ nm, compared with $5$ nm
AuNPs and $10$ nm AuNPs.

Figure~\ref{fig:2}C shows the total extinction ($\sigma_E$ =
scattering + absorption) of single AuNP normalized by the respective
scattering albedo($\bar{\omega}_0$) .  The extinction ($\sigma_E$)
normalized by the scattering albedo ($\bar{\omega}_0$) reveals that
much of the power is extinguished by the shell with $20$ nm AuNPs due
to scattering. In contrast, in the case of a shell with $5$ nm AuNPs,
power is extinguished due to absorption.

These results highlight the main mechanism of this model -- the
scattering spectra by the plasmonic core-shell assembly is modified
(scattering suppression and enhancement) due to the strong multiple
scattering by AuNPs in the shell and with the dielectric core. In the
rest of the simulation, we show results of core-shell assemblies,
where the shell comprises 20 nm and 5 nm AuNPs at a constant filling
fraction of f = 0.30. This specific choice of AuNP diameters and
filling fraction compares results between a weakly interacting
plasmonic shell (5nm AuNP) and a strongly interacting shell (20 nm
AuNP) at a high filling fraction (f = 0.30).

The effective magnitude of
suppression and enhancement of scattering can be altered by varying the diameter of
the dielectric core while maintaining a constant size of AuNP (20 nm)
and the filling fraction (f = 0.30) in the shell. Figure 3 shows a map
of $\Delta \sigma_E = \sigma_E^{CS} - \sigma_E^C$ over the visible
spectrum for various sizes of plasmonic core-shell assemblies. The
variations in the dielectric core diameter, ranging from 450 nm to 750
nm. A 750 nm core-shell assembly shows scattering suppression up to
600 nm of the visible spectrum, enhancing scattering from 600 nm to
750 nm.

\begin{figure}[htb]
    \centering
    \includegraphics[width = 6.5 cm]{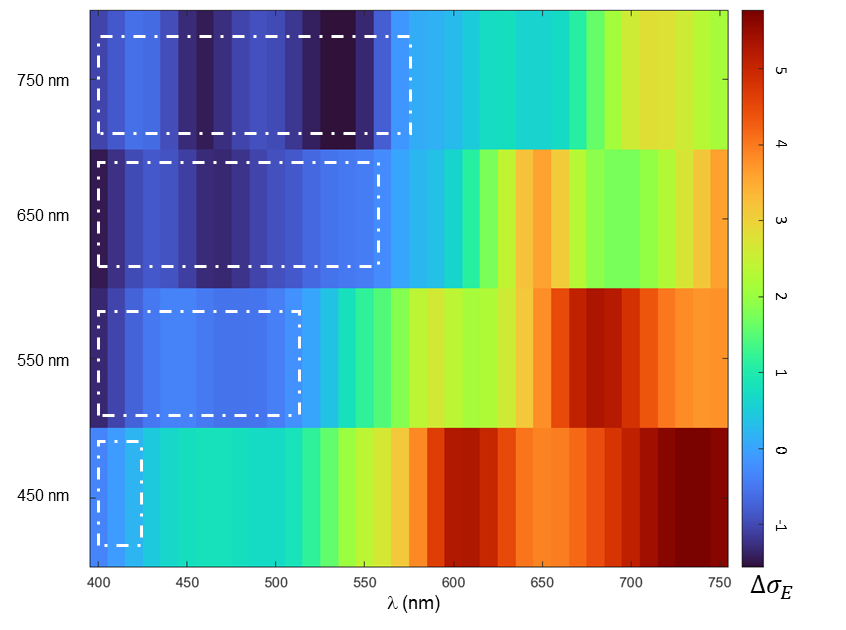}
    \caption{Comparison of $\Delta \sigma_E $
      ($\sigma_E^{CS}-\sigma_E^C$) for varied core sizes (750 nm,
      650nm, 550nm, and 450 nm) coated with 20 nm AuNPs at a filling
      fraction of 0.30. The Y-axis denotes the core diameters(nm) and
      X-axis denoting the incident wavelength(nm). The color gradient
      on the plot surface represents the magnitude of
      $\Delta \sigma_E $. A positive $\Delta \sigma_E $ refers to
      scattering enhancement and negative $\Delta \sigma_E $ is for
      scattering suppression. The region over which scattering is
      being suppressed ($\Delta \sigma_E < 0$) is highlighted by a
      white rectangle (dash-dot-dash) for each core sizes. Region
      beyond the white rectangle show scattering enhancement
      ($\Delta \sigma >0$) . }
    \label{fig: Fig_3}
\end{figure}

The assemblies of the $650$ and $550$ nm core-shell show scattering
suppression of approximately up to $550$ nm of incident light, but a
scattering enhancement beyond $550$ nm. The magnitude of the
scattering suppression is lower, and enhancement is higher by a $550$
nm core-shell assembly than the $650$ nm and $750$ nm core-shell
assemblies. A $450$ nm core-shell assembly suppresses scattering over
a very narrow wavelength range ($400$ nm - $420$ nm) but enhances
scattering over a broad visible spectrum ($430$ nm - $750$ nm). This
map emphasizes the potential that on-demand scattering suppression and
enhancement for a broad wavelength range can be achieved by varying
the diameter of the plasmonic core-shell assembly.

Plasmonic core-shell assemblies achieve total scattering suppression
and enhancement by leveraging strong multiple interactions (scattering
and absorption) between the plasmonic nanoparticles in the shell and
the dielectric core. These interactions significantly alter the
spatial and spectral profile of the scattered power. Due to this
multiple interactions of the incident light by the AuNPs, we expect
an overall decrease in coherence in the scattered field . As
a result, the angular side lobes, that are characteristic of
diffraction by a bare core, are shifted and suppressed. In other words,
we find that differential scattering cross-section ($\sigma_d$) is
smoother for the coated cores compared with the bare one, and the
smoothing is enhanced as $\sigma_E$ by an individual AuNP
increases. We study the far-field diffraction pattern of these
plasmonic assemblies and characterize them in terms of scattering
albedo ($\bar{\omega}_0$) and anisotropy ($g$) over the visible
spectrum.

\begin{figure}[htb]
    \centering
    \includegraphics[width = 6.5 cm]{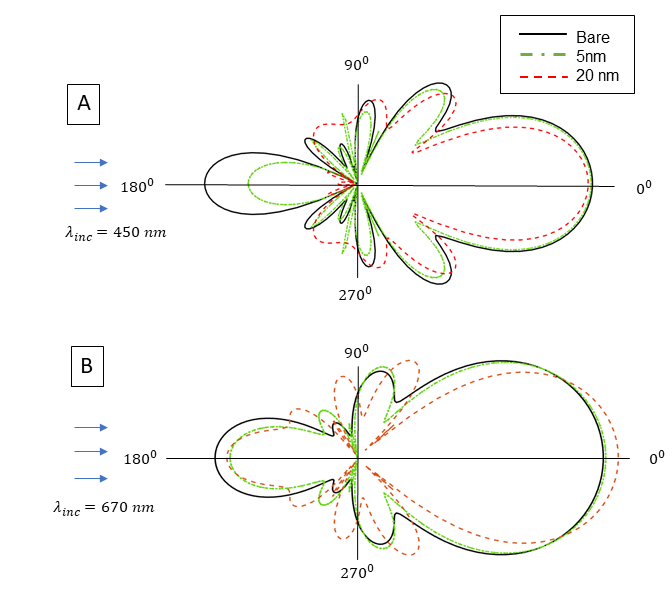}
    \caption{Polar plot comparing the far-field differential
      scattering patterns of a $750$ nm bare dielectric core and
      plasmonic core-shell assembly comprised of the same core size
      but with two types of AuNP shells ($5$ nm and $20$ nm). Incident
      light is portrayed by three horizontal arrows from left to
      right. A solid black line denotes the scattering pattern of the
      bare dielectric sphere (Bare). A green dash-dot line represents
      the $5$ nm AuNP core-shell scattering pattern ($5$ nm) and a
      $20$ nm core-shell assembly ($20$ nm) by a red dashed
      line. Filling fraction is constant for both core-shell types,
      $f = 0.30$.  (A). Scattering patterns at $450$ nm incident light
      ($\lambda_{inc} = 450$ nm). Both $5$ nm AuNPs and $20$ nm AuNPs
      shells suppress scattering at this wavelength.  (B). Comparing
      scattering patterns at $\lambda_{inc} = 670$ nm. Both plasmonic
      assemblies show scattering enhancement at this wavelength
      compared with the bare core.}
    \label{fig:Fig_4}
\end{figure}

Figure 4. Highlights and compares our findings on the $\sigma_d$ of
the 750 nm bare dielectric core (bare, solid black line), same core
(d = 750 nm) coated with 5 nm AuNP (5nm, green dash-dotted line) and
with 20 nm AuNP (20 nm, red dash line). The 20 nm and 5 nm core-shell
structures has a filling fraction f = 0.30. We compare the far field
$\sigma_d$ of these three structures at two different incident
wavelengths where the core-shell structures show suppression
and enhancement of scattering at $\lambda_{inc} =450 nm$ (Fig. 4. A) and
$\lambda_{inc} = 670 nm $ (Fig.4.B) respectively. The core-shell
structure of the 20 nm AuNP shows minimal back lobe when scattering is suppressed (Fig. 4.A), as well as reduced width and magnitude of the
main lobe. Additionally, this structure displays fewer angular side
lobes at this wavelength, resulting in a smoother appearance compared
to structures with a 5 nm core and no shell. On the contrary, the main
lobe of the 5 nm AuNP structure closely follows the main lobe of the
bare core while suppressing scattering . The back lobe of the 5 nm
AuNP structure is significantly reduced compared to the bare
dielectric sphere. Both core-shell structures show an enhancement in scattering at $\lambda_{inc} = 670 nm$(see Fig. 2A). The core-shell
structure of 20nm AuNP generates an enhanced central lobe compared
to the 5nm AuNP and bare dielectric core (Fig.4.B). The number of angular
sidelobes is also higher than the 5 nm AuNP structure and the bare core,
while scattering is enhanced. The main lobe of the 5 nm AuNP closely
follows the main lobe of the bare core. The back lobe of both
plasmonic structures has almost overlapping magnitude while enhancing
scattering(Fig.4.B). We see that a plasmonic core-shell assembly with
a 20 nm AuNP shell can manipulate the spatial and spectral profile of
the differential scattering cross section more than an assembly
composed of a 5 nm AuNp shell with the same core diameter. Furthermore, these results also suggest that plasmonic
core-shell structures emit more power in the forward (along $0^0$ deg)
than backward (along $180^{0}$ deg), regardless of whether
they suppress or enhance scattering.

We compute the scattering albedo ($\bar{\omega}_0$) of the core-shell
assemblies to determine the action of the specific extinction mechanism (scattering and absorption) on the suppression and enhancement of the general scattering.

\begin{figure}[htb]
    \centering
    \includegraphics[width = 8 cm]{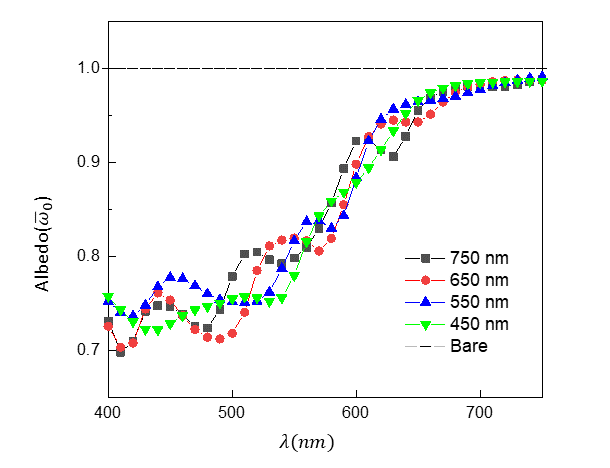}
    \caption{The scattering albedo ($\bar{\omega}_0$) of plasmonic
      core-shell structures across the visible wavelength spectrum
      shown here. The core diameters vary from 450 nm (green inverted
      triangle) to 550 nm (blue triangle), 650 nm (red filled circle),
      and 750 nm (black square). The plasmonic shells comprise 20 nm
      AuNP in all core-shell structures with a constant filling
      fraction of 0.30. The bare dielectric core has negligible
      absorption over the visible wavelength range. Consequently, the
      extinction of the incident light occurs dominantly due to
      scattering. The albedo of the bare dielectric sphere is
      indicated by a dashed horizontal line at y = 1. }
    \label{fig: fig_5}
\end{figure}

In Figure 5, we plotted the scattering albedo of the plasmonic
metastructures composed of 20 nm AuNP shell (f = 0.30) with core sizes
varied from 450 nm (green inverted triangle) to 550 nm (blue
triangle), 650 nm (red filled circle), and 750 nm (black square). The
scattering albedo of the bare dielectric core is shown with a
horizontal broken line at y = 1, as it is barely absorbing over the
visible spectrum. Almost all of the incident light is extinguished by
the bare dielectric core is due to scattering.  Albedo of all the
plasmonic core-shell structures falls between 0.7 to 0.8 for shorter
incident wavelengths ranging from $\lambda_{inc} = 400 $ to $550$
nm. The scattering albedo of the 450nm plasmonic structure changes
somewhat monotonically as we move from shorter to longer
wavelengths. Over the range of $\lambda_{inc}>= 550 nm$, the albedo
increases to a value of $\bar{\omega}_0\approx 1$ . For plasmonic
structures larger than 450 nm (i.e., 550 nm, 650 nm, and 750 nm), we
observe an oscillatory pattern in the albedo for $\lambda_{inc}$
ranging from 400 nm to 550 nm. The oscillatory pattern for the 550 nm
and 650 nm plasmonic structures becomes less pronounced beyond the
wavelength of 600 nm, while the 750 nm structures still show
noticeable oscillations up to the wavelength of 650 nm. These
oscillatory patterns reflect complex strong interactions between the
plasmonic shell and the surface of the spherical
dielectric core. The 750 nm core has a larger
surface area compared to the 450 nm core. As a result, the structure
with a 750 nm core can accommodate more gold nanoparticles
into its shell at f = 0.30, leading to a stronger interaction between
the shell and the dielectric surface. Over this shorter wavelength
range($\lambda_{inc} = 450 nm$ to $\lambda_{inc} = 550 nm$), extinction cross-section($\sigma_E$) of an individual 20 nm
AuNP is also higher (See fig 2.B), and it gradually approaches zero
beyond 550 nm incident light. Between the 450 nm and 550 nm incident
wavelength range, 750 nm, 650 nm, and 550 nm core-shell structures
show scattering suppression, while the 450 nm structures show
scattering suppression from 400 nm to 420 nm (see Fig.3). These result
suggests that approximately $70\%$ of the incident energy is extinct
by the core-shell due to scattering, and $30\%$ is due to absorption
while suppressing scattering . This is confirming that the main
mechanism of this model, which is multiple scattering between the
AuNPs and the dielectric core. Scattering suppression of the plasmonic
shell depends strongly on scattering and absorption by each individual
AuNP. When scattering by an individual metal NP is strong, power
incident on the shell of AuNPs undergoes strong multiple scattering in
the shell. When absorption by each AuNP is also strong, this strong
multiple scattering effectively yields higher absorption of the
overall power. Thus, strong scattering creates multiple interactions
with strong absorbing AuNPs thereby yielding a suppression in power
scattered by the plasmonic shell. The extinction by individual 20 nm
AuNP beyond $\lambda_{inc} = 550$ nm exponentially decreases to zero
and the scattering is the dominant extinction mechanism over this
wavelength range.  As a result, beyond 500 nm incident light,
scattering albedo increases and approaches unity, referring to all of
the incident energy over this wavelength range extinguished due to
scattering only, resulting in overall scattering enhancement by these
plasmonic metastructures.

We compute the anisotropy ($g$) to identify the amount of power that
retains its forward flow after interacting with a bare dielectric core
and the corresponding core-shell metal structures. The anisotropy
values of the metastructures are obtained by fitting the well-known
Heney-Greenstein(HG) scattering model. According to this model
$\sigma_d$ defined as,
\begin{equation}
    \sigma_d^{HG}(\theta) =
    \frac{\sigma_s}{4\pi}\frac{1-g^2}{(1+g^2-2g\cos\theta)^{3/2}}
  \end{equation}
Here, $\sigma_s$ and $g$ are the only free parameters. HG has been
extensively used to study multiple scattering by particles in
radiative transfer theory ~\cite{henyey1941}. In fact, it is often used as a
simplified model for a dielectric sphere.  Because plasmonic
core-shell structures have smoother $\sigma_d$ than the bare core, we
find HG is an appropriate model to use here. Figure 6 shows this fit
to a 750 nm core-shell structures (black dashed line) with a shell
comprising 20 nm AuNP and filling fraction varied from f = 0.3 to f =
0.05.  The differential scattering pattern of the bare dielectric
sphere (grey solid line) and the low filling fraction (f = 0.05)
plasmonic core-shell structures (blue solid line) are comparable in
terms of amplitude and phase. However, the higher filling fraction (f
= 0.30) plasmonic structure shows suppressed angular sidelobes and
smoother scattering patterns than the other two structures. We use
this fitted value for g to evaluate the difference in the angular
distribution of the scattered power by the metastructures whose core
size and AuNP size are varied and compared with the respective bare
core.

\begin{figure}[htb]
    \centering
    \includegraphics[width = 6.5 cm]{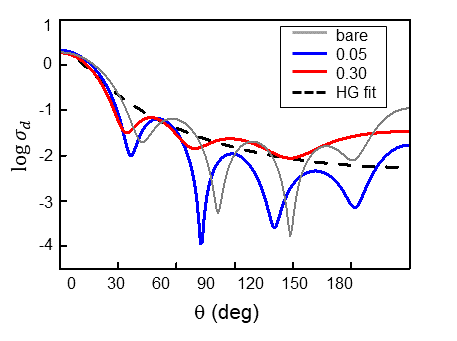}
    \caption{Comparing the differential scattering cross-section of a
      750 nm bare dielectric sphere (shown as a grey solid line) with
      the same sphere coated with 20 nm AuNP but with varying filling
      fractions of f = 0.05 (shown as a solid blue line) and f = 0.30
      (shown as a solid red line). A Henyey-Greenstein (HG) fit of the
      higher filling fraction (0.30) structure is represented by a
      black dashed line. }
    \label{fig:Fig_6}
\end{figure}

Figure 7 summarizes our findings on the anisotropy of the bare dielectric core and the corresponding plasmonic metastructures. In Fig.
7. A, we plot anisotropy of a 750 nm bare dielectric sphere ( bare,
black square), bare core coated with 5 nm AuNP ( 5nm, green inverted
triangle), and bare core coated with 20 nm AuNP ( 20 nm, red dots)
over the visible spectrum. Both core-shell structures have f = 0.30. A 5 nm AuNP shell has slightly higher anisotropy than the bare core, but a
20 nm AuNP shell can increase the anisotropy significantly. The 5 nm
AuNP shell's $g$ values closely follow the bare cores'. In the case of
20 nm AuNP shells, g values start near the bare core for shorter
wavelengths (400 nm – 500nm) but increase as the wavelength gets
longer ( $\lambda_{inc} >500 nm$).  Beyond 500nm, $\sigma_E$ of an
individual 20 nm AuNP gradually decreases (see Fig 1. B), and the
scattering governs total extinction (see Fig 1. C). These results
indicate that strong multiple scattering between the 20 nm 
plasmonic shell and the dielectric core promotes a higher forward
directionality of incident power compared to a 5 nm AuNP shell.

\begin{figure}[htb]
    \centering
    \includegraphics[width = 7.0 cm]{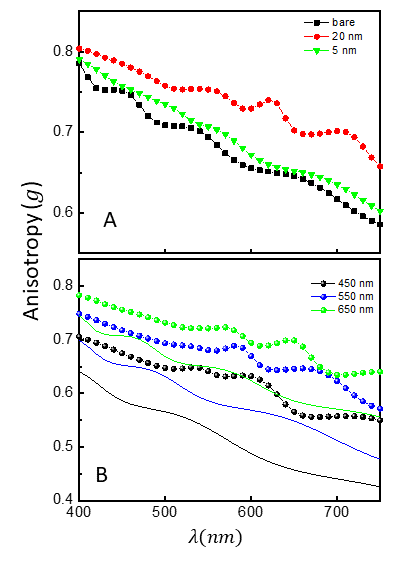}
    \caption{(A) Comparing anisotropy ($g$) of a 750 nm bare
      dielectric core ( bare , black square) with the plasmonic
      core-shell structure over the visible spectrum. Plasmonic
      metastructures are composed of a core of same diameter but
      coated with 5 nm ( 5nm, green triangle) and 20 nm AuNPs (20 nm,
      red dots) as shell. AuNP filling fractions is 0.30 for both
      shells.  (B) Anisotropy of bare cores with varied diameters are
      compared with the corresponding plasmonic core-shell
      structures. Plasmonic shells are comprised of 20 nm AuNP for all
      core diameters and filling fraction is 0.30. Core diameters are
      varied from 450 nm(black) to 550nm (blue) , and 650
      nm(green). Anisotropy of the bare cores are depicted using a
      solid lines and the respective core-shells are using dotted
      lines with same colors.  }
    \label{fig:fig_7}
\end{figure} 

In Figure 7B, we compared the anisotropy of various core diameters
with the corresponding plasmonic core-shell metastructures across the
visible spectrum. The core diameters ranged from 450 nm (black) to 550
nm (blue) and 650 nm (green). The anisotropy of the core-shell
structures is represented using dotted spheres, while the respective
bare cores are depicted using solid lines of the same color. The
plasmonic shell consists of 20 nm AuNP at f = 0.30 for all core-shell
structures. Coating a bare dielectric sphere with randomly distributed
20 nm AuNPs increases the anisotropy of all the core sizes studied
here compared to their corresponding cores (Figure 7B). A 450 nm
core-shell structure enhances anisotropy more than the 650 nm and 550
nm core-shell assemblies. These results indicate that plasmonic metastructures increase anisotropy for a wide range of core
sizes across the visible spectrum. This increase in anisotropy occurs
independently of whether a specific core-shell structure suppresses or
enhances scattering for a particular incident wavelength. In other
words, plasmonic AuNP coating results in a preferential concentration
of the scattered power in the forward direction across the visible
spectrum.

\section{Conclusions}
\label{sec:conclusions}

Our computational model demonstrates that a coat of randomly
distributed AuNPs on a dielectric core can be optimized to modulate
the mesoscopic optical properties in the visible spectrum in more than
one way. A coating consisting of $20\, \text{nm}$ AuNPs with moderate
filling fractions $0.2 < f < 0.3$ results in significant scattering
suppression up to $\lambda = 600\, \text{nm}$ for cores larger than
the incident wavelength. But a slight variation of $f$, between $0.1$
-- $0.2$, leads to scattering enhancement in the spectral regime
$\lambda > 650\, \text{nm}$ for the same coating. This substantial
difference in optical response highlights the versatility of this
platform. In addition to this spectral modulation, we observe that the
presence of the cover results in a preferential concentration of the
scattered power in the forward direction. This reshaping of the
angular distribution of power occurs when scattering is suppressed and
enhanced. Our results further underscore the critical role of
absorption versus scattering by the AuNPs, as we establish that some
absorption by the AuNP cover is needed to suppress the scattering of
the core. However, scattering is suppressed most when $f$ is optimal
to promote both multiple scattering and absorption with a stronger
dependence on former. It is for this reason that $20\, \text{nm}$
AuNPs yield more suppression than $5\, \text{nm}$ AuNPs. In fact, it
is because of multiple scattering in the cover that the angular
distribution of scattered power by the coated cores is qualitatively
different from that by the bare core. Multiple scattering decreases
the overall coherence of scattered light and consequently, suppresses
the angular sidelobes in the distribution of scattered power. As a
result the forward peak in scattering is more pronounced and the
anisotropy factor increases. Through this investigation of the
mesoscopic optical properties of nano-assembled plasmonic coatings, we
have identified parameter regimes where the cover of AuNPs lead to
scattering suppression or enhancement. These results can then guide
decisions on tuning AuNP size and filling fraction for a wide range of
optical and photonic applications in the visible spectrum, ranging
from near-field microscopy to high-resolution imaging.
\section*{Disclosures}
The authors declare no conflicts of interest.

\bibliography{KhanGhoshKim-2024}

\end{document}